\documentclass[10pt,twocolumn,floatfix,prl,superscriptaddress]{revtex4-1}
\usepackage{amsmath,amssymb,amsthm,mathrsfs,amsfonts,dsfont,amstext} 
\usepackage{textcomp,pbox}
\usepackage[export]{adjustbox}
\usepackage{bm}
\usepackage{dcolumn,booktabs,url}
\usepackage[scaled]{helvet}
\usepackage{sansmath,gensymb}
\usepackage{tikz,graphicx,transparent,color}
\usepackage{multirow}
\usepackage[separate-uncertainty = true]{siunitx}
\usepackage{comment}
\usepackage{physics}
\usepackage{pdfpages}
\usepackage{array}

\makeatletter
\AtBeginDocument{\let\LS@rot\@undefined}
\makeatother

\newcolumntype{C}[1]{>{\centering\let\newline\\\arraybackslash\hspace{0pt}}m{#1}}

\usepackage[colorlinks=true]{hyperref}
\usepackage{graphicx}

\hypersetup{
     colorlinks   = true,
     citecolor    = red,
     linkcolor    = blue,
     urlcolor     = red     
}

\graphicspath{{./figs/}}

\newcommand{\TS}{$\SI{21.4(7)}{ns}$}

\newcommand{\ketup}{{\ket{\uparrow}}}
\newcommand{\ketdown}{{\ket{\downarrow}}}
\newcommand{\ketUp}{{\ket{\Uparrow}}}
\newcommand{\ketDown}{{\ket{\Downarrow}}}

\newcommand{\braup}{{\bra{\uparrow}}}
\newcommand{\bradown}{{\bra{\downarrow}}}

\begin{document}

\newcommand{\TitleName}{A coherent spin-photon interface with waveguide induced cycling transitions}
\title{\TitleName}

\newcommand{\AffCPH}{Center for Hybrid Quantum Networks (Hy-Q), The Niels Bohr Institute, University~of~Copenhagen,  DK-2100  Copenhagen~{\O}, Denmark}
\newcommand{\AffBasel}{Department of Physics, University of Basel, Klingelbergstra\ss e 82, CH-4056 Basel, Switzerland}
\newcommand{\AffBochum}{Lehrstuhl f\"ur Angewandte Fest\"orperphysik, Ruhr-Universit\"at Bochum, Universit\"atsstra\ss e 150, 44801 Bochum, Germany}

\author{Martin Hayhurst Appel}
\affiliation{\AffCPH{}}
\author{Alexey Tiranov}
\affiliation{\AffCPH{}}

\author{Alisa Javadi}
\affiliation{\AffBasel{}}
\author{Matthias C. L\"obl}
\affiliation{\AffBasel{}}

\author{Ying Wang}
\affiliation{\AffCPH{}}

\author{Sven Scholz}
\affiliation{\AffBochum{}}
\author{Andreas D. Wieck}
\affiliation{\AffBochum{}}
\author{Arne Ludwig}
\affiliation{\AffBochum{}}

\author{Richard J. Warburton}
\affiliation{\AffBasel{}}
\author{Peter Lodahl}
\affiliation{\AffCPH{}}

\email[Email to: ]{d.authors@nbi.ku.dk}

\date{\today}

\begin{abstract}
Solid-state quantum dots are promising candidates for efficient light-matter interfaces connecting internal spin degrees of freedom to the states of emitted photons. However, selection rules prevent the combination of efficient spin control and optical cyclicity in this platform. By utilizing a photonic crystal waveguide we here experimentally demonstrate optical cyclicity up to $\approx15$ through photonic state engineering while achieving high fidelity spin initialization and coherent optical spin control. These capabilities pave the way towards scalable multi-photon entanglement generation and on-chip spin-photon gates.
\end{abstract}

\maketitle 


Single solid-state spins play an important role in modern quantum information technologies \cite{Degen2017,Awschalom2018,Atature2018}. A key resource is a coherent light-matter interface connecting spins and photons compatible with high fidelity spin manipulation and long distance distribution of quantum states.
Self-assembled semiconductor quantum dots (QDs) are currently considered one of the most promising systems thanks to their high photon generation rate, good optical and spin coherence properties, and efficient integration into photonic nanostructures~\cite{Lodahl2015}. The latter is important for realizing strong light-matter interaction and high collection efficiencies \cite{Ding2016,Uppu2020}. 

An application particularly well suited to QDs is the deterministic generation of multiphoton entangled states such as photonic cluster states~\cite{Lindner2009}. These maximally entangled states have applications in measurement-based quantum computing~\cite{Raussendorf2001} and quantum repeaters~\cite{Borregaard2019,Azuma:2015ab} and so far one-dimensional cluster states containing 3 qubits have been generated with QDs~\cite{Schwartz2016}. A recent protocol based on time-bin encoded photonic qubits has been put forward allowing scaling up to tens of high-fidelity entangled photons for experimentally relevant parameters~\cite{Lee2019,Tiurev2020a}. Crucially, this protocol requires an optical transition with high cyclicity, i.e. a transition where the excited state decays selectively to one of the two ground states thereby preserving the spin.

\begin{figure}[t]%
\begin{center}
\includegraphics[width=1\columnwidth]{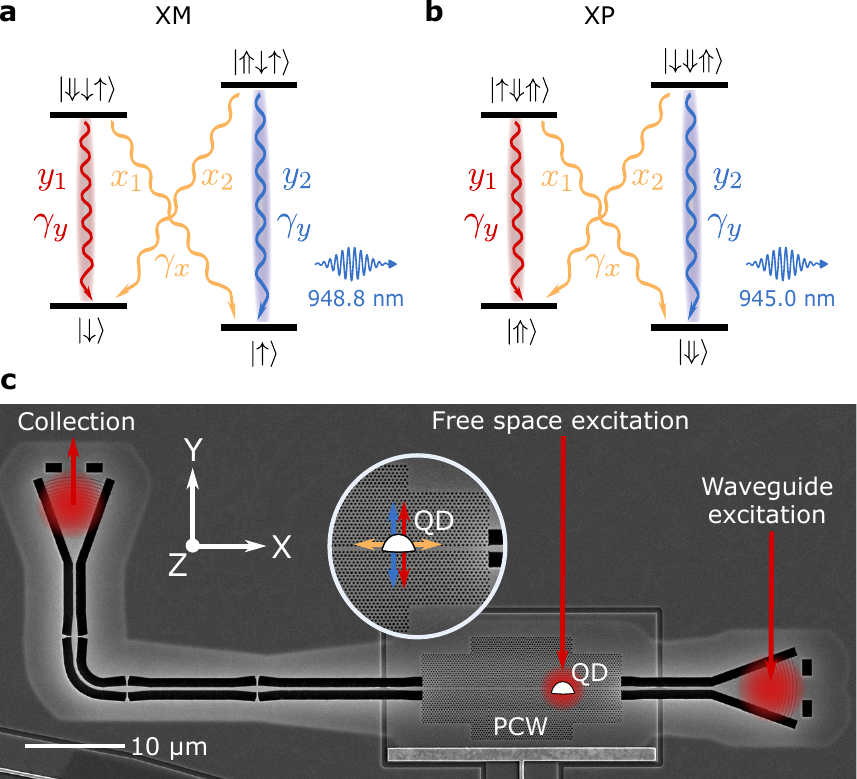}
\end{center}
    \caption{(color online). \textbf{(a,b)} Energy level diagram of a (a) negatively charged and (b) positively charged QD in an in-plane magnetic field. $\uparrow,\downarrow$ indicate electron spins and $\Uparrow,\Downarrow$ indicate hole spins. The four linear dipoles have equal decay rates in bulk but are selectively enhanced by the waveguide. \textbf{(c)} Scanning electron microscope image of the two sided waveguide containing a PCW section, QD and grating couplers. Circular insert shows ideal orientation of linear dipoles for selective waveguide coupling. }
    \label{fig:1}%
\end{figure}

In the context of QDs, the cyclicity corresponds to the ratio between the decay rates of the \textit{vertical} transitions and the \textit{diagonal} transitions, see Fig. \ref{fig:1}a. High cyclicity can be achieved by operating the QD in an out-of-plane magnetic field (Faraday geometry) where the diagonal transitions are only weakly allowed~\cite{Dreiser2008}. This configuration, however, obstructs fast all-optical spin control. 
For this reason, an in-plane magnetic field (Voigt geometry) has been indispensable for achieving spin control as demonstrated in bulk media~\cite{Press2008,Gangloff2019,Bodey2019}, cavities~\cite{Carter2013,Sun2016,Wang2019a} and nanobeam waveguides~\cite{Ding2019}. 
In this geometry, vertical and diagonal transitions posses equal magnitude precluding cycling transitions. However, cyclicity may be induced by a nanostructure through selective enhancement of the optical transitions. Such an enhancement has recently been demonstrated in cavity systems including single rare-earth ion spins~\cite{Raha2020}. QDs coupled to photonic crystal cavities~\cite{Carter2013,Sun2016} and  micropillar cavities~\cite{Lee2019,Wang2019a} also exhibit selective enhancement, although the cyclicity has not been explicitly demonstrated. 

Here, we report the first realization of QD cycling transitions in the Voigt geometry due to the selective enhancement provided by a photonic crystal waveguide (PCW). 
The mechanism is fundamentally broadband as it relies on the orthogonally polarized dipoles having vastly different projections along the local electric field, as quantified by the projected local density of optical states in the PCW~\cite{Lodahl2015}.
This approach has several advantages as it allows high cooperativity of multiple non-degenerate optical transitions, does not require substantial energy splittings of the optical transitions, and allows direct integration into photonic circuits. 
We measure a cyclicity of 11.6 (14.7) for the negative(positive) charge states of the same QD despite a 3.8~nm spectral separation. 
 In addition we achieve 98.6\% spin preparation fidelity, $T_2^* = $~\TS{} spin dephasing time, and all-optical spin control on the positively charged QD. These achievements together with the efficient photon collection of the PCW make our platform highly advantageous for the generation of multiphoton entangled states and the implementation of deterministic spin-photon quantum gate operations~\cite{Borregaard2019,Witthaut2012}.

The nanostructure under consideration is a PCW made from a suspended GaAs membrane and connected to grating couplers at either ends, see Fig. \ref{fig:1}c. The membrane comprises a p-i-n diode grown in the $z$-direction with the intrinsic layer containing self assembled InAs QDs~\cite{Lobl2019}. Applying a forward bias voltage deterministically charges the QD with a single electron (XM configuration) whereas additional optical induction allows the creation of a metastable hole state (XP configuration) with $>\SI{16}{\micro s}$ lifetime (see supplementary material).

Applying an in-plane magnetic field results in a four level spin system for the XM (Fig. \ref{fig:1}a) and XP (Fig. \ref{fig:1}b). The XM~(XP) systems comprise two Zeeman split ground states with a single electron~(hole) spin and two Zeeman split trion states containing a magnetically active hole~(electron) and a singlet of two electrons~(holes). The selection rules, which are functionally identical for XM and XP, result in two $\Lambda$-systems, each containing an $x$- and a $y$-polarized linear dipole. While the associated radiative decay rates $\gamma_x$ and $\gamma_y$ are identical in bulk~\cite{Warburton2013} the photonic environment of the PCW may selectively enhance and suppress the orthogonally polarized transitions, yielding a cyclicity $C=\gamma_y/\gamma_x \geq 1$.  The decay rates can be further decomposed into a waveguide (\textit{wg}) and radiative (\textit{rad}) component, $\gamma_i=\gamma_{i,wg}+\gamma_{i,rad}$, $i=\{x,y\}$. $\gamma_{x,rad}$ and $\gamma_{y,rad}$ are determined by the coupling to a continuum, are similar in magnitude and generally strongly suppressed throughout the PCW~\cite{Javadi2018:OSAB}. In contrast, $\gamma_{x,wg}$ and $\gamma_{y,wg}$ are determined by the projected coupling onto the single polarized waveguide mode and can thus vary between zero and a highly enhanced rate given by the high optical density of states. This is the origin of the high cyclicity in the PCW which can be sensitively controlled by the position of the QD~\cite{Javadi2018:OSAB}. 

\begin{figure}[h]%
\includegraphics[width=1\columnwidth]{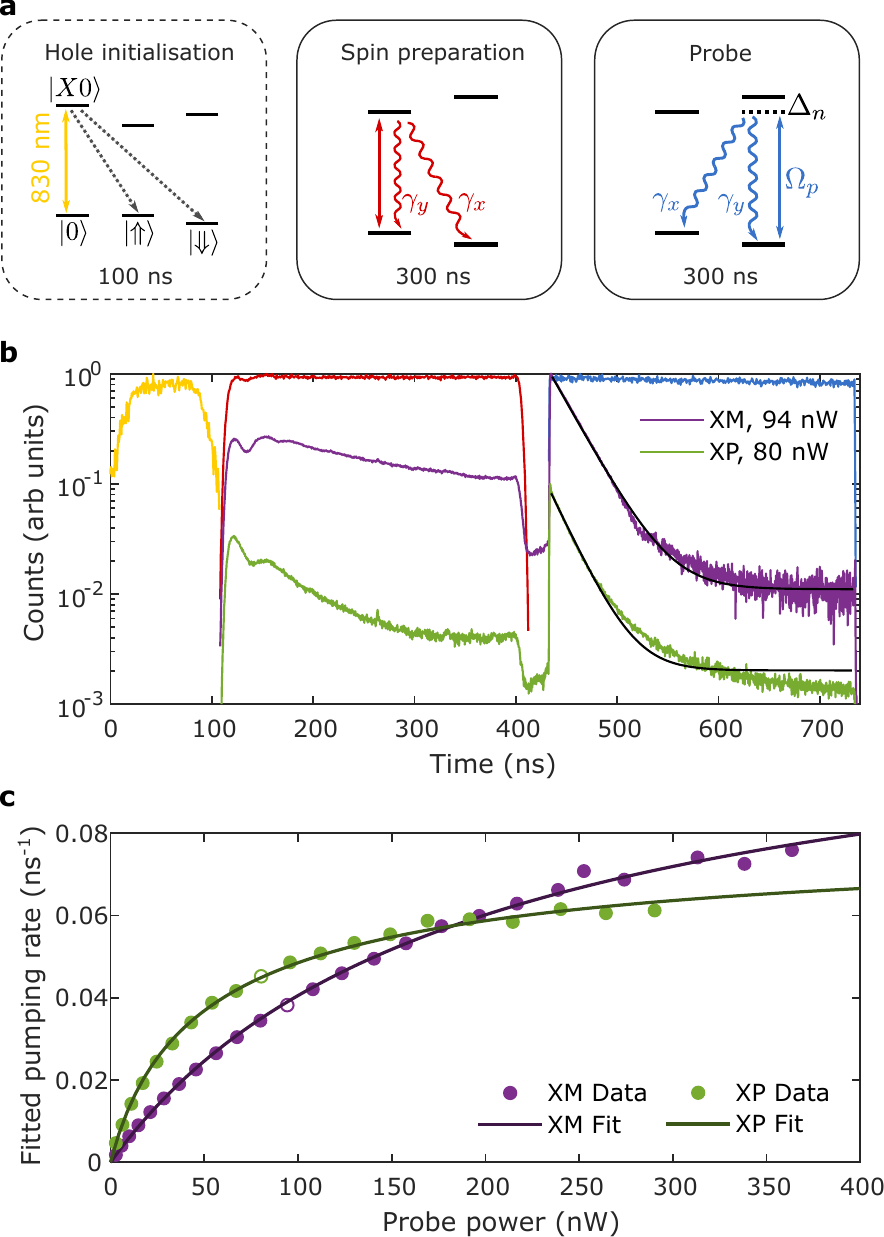}
    \caption{(color online). \textbf{(a)} Pulse sequence with two resonant pumping pulses and an additional photocreation pulse used to populate the hole. Level structures correspond to Fig.~\ref{fig:1}b and Fig.~\ref{fig:1}c. \textbf{(b)} Spin pumping fluorescence histograms taken at $B_y=\SI{2}{T}$. Yellow, red and blue traces show laser pulse shapes while purple~(green) traces show fluorescence from XM~(XP) experiments with laser background subtracted. The green curve is shifted by $\times 0.1$ for clarity. Black lines indicate fits with the model $I(t)=I_0+I_1e^{-\gamma_{osp}t}$. \textbf{(c)} Extracted pumping rates during the probe pulse as a function of probe power for XP and XM. Saturation fits follow Eq.~\eqref{eq:gamma_osp} and only contain $\gamma_x$ and $P_{sat}$ as free parameters. Unfilled dots correspond to traces in (b).
    }\label{fig:2}%
\end{figure}

We now determine the cyclicity of the XM and XP systems with the two-color pump/probe pulse sequence in Fig.~\ref{fig:2}a. For XM, a narrow band laser pulse with fixed power prepares $\ketup$ by driving $\ketdown\rightarrow\ket{\Downarrow\downarrow\uparrow}$. The probe pulse performs the opposite operation, driving $\ketup\rightarrow\ket{\Uparrow\downarrow\uparrow}$ and preparing $\ketdown$. Both pulses are $y$-polarized to minimize coupling to the $x$-transitions. In the limit of $C\gg1$ the rate of optical spin pumping $\gamma_{osp}$ saturates according to
\begin{align}
 \gamma_{osp} = \gamma_x\int_{-\infty}^{\infty} \frac{\Omega_p^2}{2\Omega_p^2+\gamma_0^2+4\Delta_n^2}\mathcal{N}(\Delta_n;\sigma)d\Delta_n\;,\label{eq:gamma_osp}
\end{align}
where $\gamma_0=\gamma_x+\gamma_y$ is the excited state lifetime, $\Omega_p=\gamma_0\sqrt{P/P_{sat}}$ is the probe Rabi frequency, $P$ is the optical power and $P_{sat}$ is the saturation power. For completeness, we include slow spectral diffusion via the detuning $\Delta_n$ which is drawn from a normal distribution $\mathcal{N}(\Delta_n;\sigma)$ with standard deviation $\sigma$, although the effect on the cyclicity estimate turns out to be minor ($<3$\%).  By varying the probe power and fitting the fluorescence histograms (Fig.~\ref{fig:2}b) a set of spin pumping rates are obtained. 
These rates are plotted in Fig.~\ref{fig:2}c and fitted with (\ref{eq:gamma_osp}). A characterization (supplementary material) of the XM yields $\gamma_0^{(XM)}=\SI{3.07(6)}{ns^{-1}}$ and $\sigma^{(XM)}/2\pi=\SI{140}{MHz}$. The pumping measurement then yields $\gamma_x^{(XM)}=\SI{0.243(5)}{ns^{-1}}$ and a cyclicity of $C^{(XM)}=(\gamma_0^{(XM)}-\gamma_x^{(XM)})/\gamma_x^{(XM)}=\SI{11.6(4)}{}$. 
The XP cyclicity is measured with the same method but with the inclusion of a 100~ns 830~nm laser pulse which populates~\cite{Delteil2016} the hole via the neutral exciton (X0), see Fig.~\ref{fig:2}a. Here we find $\sigma^{(XP)}/2\pi=\SI{345}{MHz}$, $\gamma_0^{(XP)}=\SI{2.48(2)}{ns^{-1}}$, $\gamma_x^{(XP)}=\SI{0.158(2)}{ns^{-1}}$  and a cyclicity of $C^{(XP)}=\SI{14.7(2)}{}$.
Hence, XM and XP both demonstrate substantially increased cyclicity owing to both an inhibited $\gamma_x$ and an enhanced $\gamma_y$ compared to the expected bulk values~\cite{Warburton2013}. 
This enhancement occurs for both XM and XP despite a spectral separation of 3.8 nm, demonstrating the broadband enhancement provided by the waveguide.  

High-fidelity spin initialization is the essential starting point for applications of spin-photon interfaces. By analysing the steady state fluorescence at the end of the spin pumping histograms we estimate lower bounds of the spin initialization fidelitities $F_s^{(XM)} = \bra{\downarrow}\rho\ket{\downarrow}=99.1\%$ and $F_s^{(XP)} = \bra{\Uparrow}\rho\ket{\Uparrow}=98.6\%$, which are the highest values so far reported in photonic nanostructures~\cite{Lee2019,Sun2016,Carter2013}. 
In contrast to cross polarization experiments, our laser polarization control allows us to avoid driving the $x$-transitions which otherwise reduce the initialization fidelity through re-pumping. The $1/e$ spin pumping time when driving a $y$-transition is limited to $2/\gamma_x^{(XP)}=\SI{12.7}{ns}$, although driving an $x$-transition would result in significantly faster initialization at the cost of reduced fidelity owing to the $x$-transition's increased frequency overlap. 

\begin{figure}[t]%
\includegraphics[width=1\columnwidth]{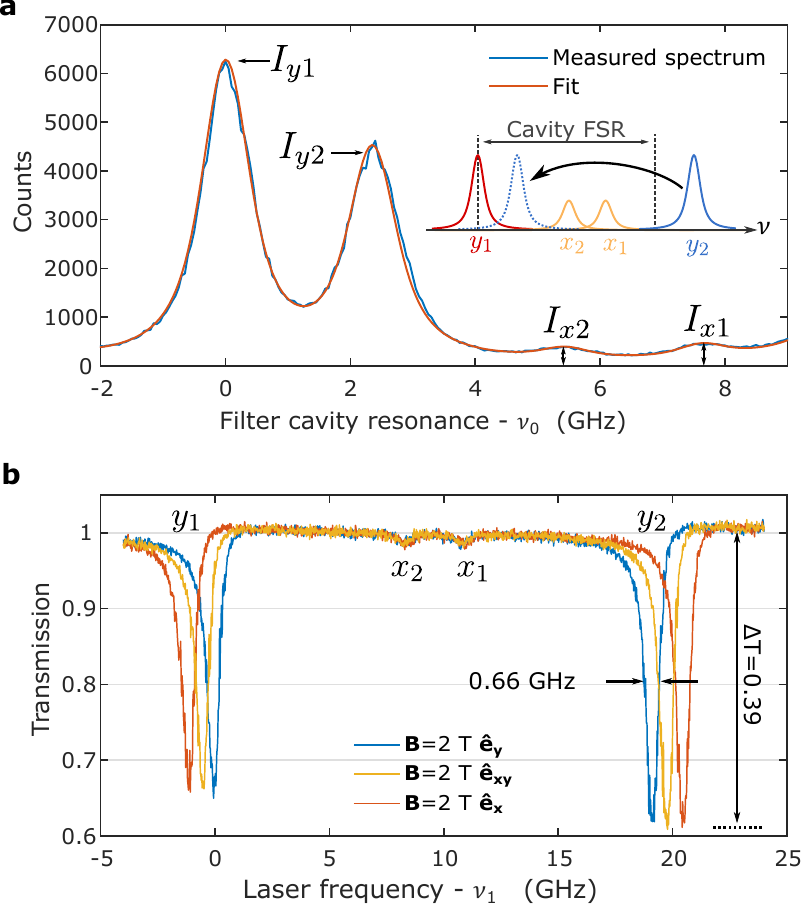}
    \caption{(color online). \textbf{(a)} Emission spectrum of the XM at $B_y=\SI{1.3}{T}$ generated by p-shell excitation and analyzed by a scanning Fabry-Perot cavity. The insert shows the re-ordering of the emission peaks due to the cavity. $\nu_0=315.97$~THz. \textbf{(b)} Normalised waveguide transmission in the presence of an XM in the cotunnelling regime. Colored lines denote different in-plane magnetic field orientations. A strong asymmetry between the dip amplitudes of the $x$ and $y$ dipoles persists for all magnetic field directions. $\nu_1=315.94$~THz.
    }
    \label{fig:3}%
\end{figure}

To further investigate the origin of the cyclicity we explicitly probe the coupling between the XM dipoles and the mode of the PCW. First, we analyze the spontaneous emission spectrum following continuous wave p-shell excitation. This method allows effective elimination of laser background (see Methods) and population of both negative trions. The emission is analyzed using a scanning Fabry-Perot filter cavity, revealing all four optical transitions, see Fig.~\ref{fig:3}a. By fitting the spectrum we extract the intensity ratios $I_{y1}/I_{x1}=21.8 \pm 0.5$ and $I_{y2}/I_{x2}=20.3 \pm 0.6$ which do not differ significantly (1.9$\sigma$ deviation) and quantify the trions' strong preference to decay into the PCW via the $y$-transitions. Taking the average, we estimate a waveguide-coupling asymmetry of $A^{(XM)}=\gamma_{y,wg}^{(XM)}/\gamma_{x,wg}^{(XM)} = 21.1\pm0.4$. It should be stressed that this measurement only probes the dipole/waveguide coupling where as the cyclicity also receives contributions from radiative modes causing it to be lower.
\begin{figure*}[t]%
\begin{center}
\includegraphics[width=1\textwidth]{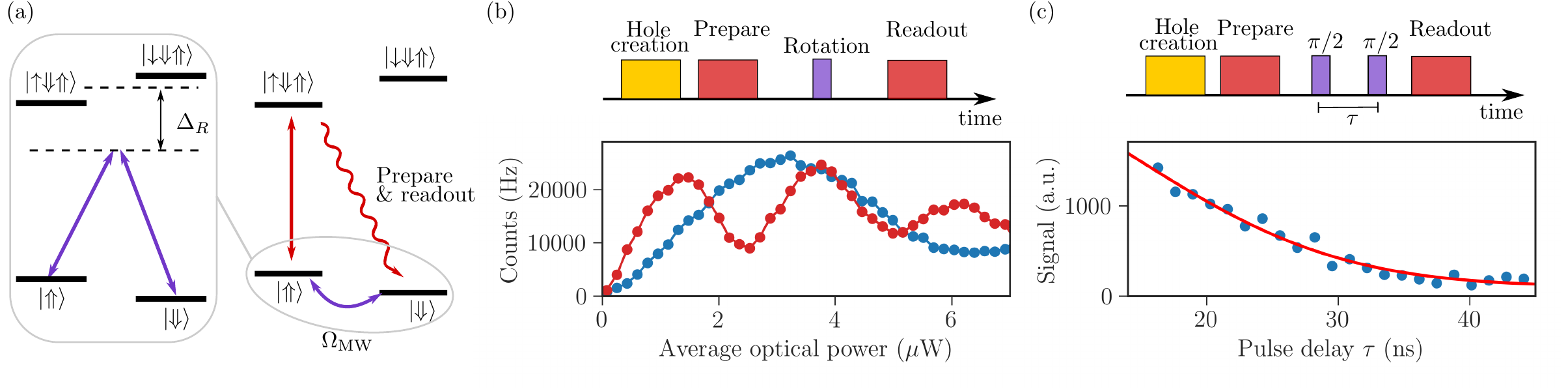}
\end{center}
    \caption{(color online)
    	Coherent spin control in a $B_y=2$~T magnetic field. \textbf{(a)} Raman energy level diagram. The detuned Raman laser (purple) results in an effective ground state coupling $\Omega_{MW}$. 
    	\textbf{(b)} Readout fluorescence of XP following a 20~ns Raman pulse with varying power and detunings $\Delta_R= \SI{290}{GHz}$ (red) and $\Delta_R= \SI{790}{GHz}$ (blue). The maximum $\Omega_{MW}/(2\pi)=\SI{150}{MHz}$ is limited by the available optical power and the weak free space coupling of the PCW.
		\textbf{(c)} Readout fluorescence of XP after applying a Ramsey sequence with two $\pi/2$ pulses optimized using the data in (b). The signal is fit using Eq.~\eqref{eq:Ramsey} resulting in an effective spin dephasing time of $T_2^* = $~\TS{}. Short pulse delays are unavailable due to equipment limitations.
		}
    \label{fig:4}%
\end{figure*}

Next, we probe the coherent interaction between the four QD dipole transitions and the waveguide mode by measuring waveguide transmission. Previous experiments on charged QDs in waveguides were conducted on XM in a Faraday configuration where single photon switching was demonstrated~\cite{Javadi2018:Spin}. In the present study we observe all four XM dipoles in the Voigt geometry. 
To avoid optical spin pumping we operate in the co-tunnelling regime where tunnelling to the back contact of the diode provides a randomization of the electron spin with the rate $\kappa_{co}/2\pi\approx 1-10$~MHz~\cite{Dreiser2008}.
In the limit $\kappa_{co}\gg\gamma_{osp}$ spin flips randomize the spin in a thermal state $\hat{\rho}\approx 0.53\ketup\braup+0.47\ketdown\bradown$ given the temperature $T=\SI{4}{K}$ and in a 2~T magnetic field. 
Fig.~\ref{fig:3}b shows the XM transmission when operating in the co-tunneling regime. We observe a 0.66 GHz linewidth (1.35$\times$ natural linewidth) and transmission dip amplitudes up to 39\% where a factor $\approx2$ reduction in the transmission dips stems from the thermal mixture. This pronounced transmission spectrum directly demonstrates the coherent nature of the spin-photon interface, which is required for applications such as single photon transistors \cite{Witthaut2010,Chang2007}, deterministic Bell state analyzers~\cite{Witthaut2012} and quantum gates \cite{Duan2004, Rosenblum2017}. Furthermore, the high asymmetry between $x$- and $y$-dipoles is directly visible, and we extract a ratio of $\approx27$ between the $y$ and $x$ dips. The detailed modelling of the transmission data is beyond the scope of this paper. Strikingly, rotating the magnetic field does not alter the dip amplitudes suggesting an insensitivity of the dipole orientations to the magnetic field. This has previously been observed in Stranski-Krastanov grown QDs and is attributed to an anisotropic QD shape or strain profile~\cite{Koudinov2004}.

It is of course instructive to consider the limiting factors of the cyclicity. In principle, one can selectively enhance $\gamma_{y,wg}$ indefinitely by approaching the PCW bandedge at which the density of states diverges~\cite{Javadi2018:OSAB}. In practice, this limit is unreachable due to fabrication imperfections~\cite{Garcia2017}. Simulations~\cite{Javadi2018:Simulations} predict $C=144$ being achievable at a group index of 56, corresponding to a realistic Purcell enhancement of about 10~\cite{Arcari2014}. Realising the maximal cyclicity requires positioning of the QD and correct orientation between the QD dipoles and the PCW mode polarization. Deterministic fabrication of a PCW relative to a QD has already been demonstrated~\cite{Pregnolato2019} so additional control over the PCW rotation is feasible.

Finally, to properly evaluate our system's usefulness for quantum information processing applications we consider quantum control of the hole spin due to its increased coherence time over the electron spin~\cite{Prechtel2016,Huthmacher2018}. We employ the Raman spin control scheme recently demonstrated in Ref.~\cite{Bodey2019} which allows flexible phase control of the spin and elaborate electronically defined pulse sequences. A circularly polarized CW laser is red detuned by $\Delta_R$ from the main optical transitions and amplitude modulated at frequency $\Delta_D/2$ resulting in two sidebands matching the ground state splitting. The optical field then creates an effective coupling between $\ketUp$ and $\ketDown$ with Rabi frequency $\Omega_{MW}\propto P_{R}/\Delta_R$ where $P_{R}$ is the Raman laser power, see Fig.~\ref{fig:4}a. We first verify the existence of Rabi oscillations between $\ketUp$ and $\ketDown$ by varying the Raman pulse power and $\Delta_R$, see Fig.~\ref{fig:4}b. Applying the Ramsey pulse sequence in Fig.~\ref{fig:4}c yields the free induction decay of the hole spin which is fitted according to Ref.~\cite{Bodey2019} 
\begin{align}
I(\tau)=I_0 e^{-(\tau/T_2^*)^2},\label{eq:Ramsey}
\end{align}
and yields a dephasing time of $T_2^*= $~\TS{}. This is long compared to the emitter lifetime, $T_2^*\gamma_0^{XP}=54$, thus allowing a considerable number of photons to be emitted within the spin dephasing time. 
This hole spin $T_2^*$ is on par with the performance found only in bulk-like samples and at higher magnetic fields~\cite{Delteil2016,Godden2012}, demonstrating the capability of overcoming deteriorating noise processes in the nanostructures despite the near-by proximity of the QD to surfaces. For comparison, previous experiments in nanostructures reported a $T_2^* = 2.11$ ns of a hole spin in micropillar cavities \cite{Lee2019} and below one nanosecond for electron spins in planar photonic cavities~\cite{Sun2016,Carter2013}.

To summarize, we have successfully demonstrated how the photonic environment of a QD can be engineered to provide broadband selective enhancement of optical transitions resulting in cycling transitions in the Voigt geometry. The PCW accomplishes this without the need for high magnetic fields and QD tunability otherwise required by a cavity. The ability to excite the QD via a side channel enables both high fidelity spin preparation and optical spin control. The demonstrated $C^{XP}\approx 15$ enables generation of multi-photon cluster states using time-bin encoding~\cite{Tiurev2020a} and would also greatly improve single-shot spin readout efficiency. The demonstration of a coherent spin-photon interface will be of immediate use for spin-photon and photon-photon quantum gates mediated by the efficient coupling between the QD and the PCW.
\\ \\
The authors thank C. Starup for building the filter cavity and R. Uppu for valuable discussions. We gratefully acknowledge financial support from Danmarks Grundforskningsfond (DNRF 139, Hy-Q Center for Hybrid
Quantum Networks), H2020 European Research Council (ERC) (SCALE), Styrelsen for Forskning og Innovation
(FI) (5072-00016B QUANTECH), the European Union’s Horizon 2020 research and innovation programme under grant agreement No. 820445 and project name Quantum Internet Alliance and
Deutsche Forschungsgemeinschaft (DFG) (TRR 160).
A.J., M.C.L., and R.J.W. acknowledge financial support from SNF Project No. 200020\_156637 and from NCCR QSIT. A.J. acknowledges support from the European Unions Horizon 2020 Research and Innovation Programme under the Marie Skłodowska-Curie grant agreement No. 840453 (HiFig).

\bibliography{PRL_Manuscript_MHA.bbl}

\newpage

\clearpage

\end{document}